\documentclass[a4paper,12pt,oneside,leqno]{article}
\usepackage[latin1]{inputenc}
\usepackage[T1]{fontenc}
\usepackage[italian]{babel}
\usepackage[mathscr]{euscript}
\usepackage{amsmath,exscale,latexsym,amssymb,amsthm,amsfonts}
\title{Some critical remarks on Landau's (macroscopic) phase transition theory\footnote{This work is part of a
communication given at the XCI-th National Congress of the Italian Physical Society (SIF), III-th section:
Astrophysics and Cosmic Physics, held to Department of Physics and Astronomy of the University of Catania,
September 26 - October 1, 2005.}}
\author{Giuseppe Iurato\footnote{e-mail: iurato@dmi.unict.it}}
\date{}
\begin{document}
\maketitle
\begin{quotation}\small \bf Abstract. \it This paper explain the existence of a particular formal
model (drew from Theoretical Astrophysics) whose thermodynamical phenomenology shows a possible second order
phase transition (according to Landau's Thermodynamical Theory) that seems does not verify the
(Birman-Goldrich-Jari\'{c}) ''chain subduction criterion'' and the (Ascher's) ''maximality criterion'' of
Landau's Phenomenological Theory. Therefore, it follows that Landau's Phenomenological Theory is more
restrictive than the Landau's Thermodynamical Theory.\end{quotation}
\section{Introduction}Lev D. Landau ([La], [LL5]) given two theoretical formulations of the (macroscopic)
second order phase transition theory, the first called \it (Landau's) macroscopic Thermodynamical Theory, \rm
and the second called \it (Landau's) macroscopic Phenomenological Theory\rm. These theories are based
(especially the second) on some symmetry principles and elementary methods of the representation theory of
groups.\\\\In this context, there exists criteria\footnote{See [As1], [As2], [AK], [Bi2], [Bo], [CLP], [GL],
[Ja1], [ITO], [LL5], [Ly], [Mi1], [Mi3], [SH], [Th], [To1].} that give necessary or sufficient conditions for a
second order phase transition (according to Landau).\\ Among these, we recall the \it
(Birman-Goldrich-Jari\'{c}) chain subduction criterion\footnote{See [Bi1], [CLP], [GB], [Ja1], [Ja2], [Ja3],
[Ja4], [Ja7], [JB], [SA], [To1].} \rm and the \it (Ascher's) maximality criterion\footnote{See [As1], [As2],
[AK], [Bi2], [CLP], [Ja1], [Ja6], [To2].}.\\\\ \rm In this paper, we compare these two criteria with a
well-known formal model of theoretical astrophysics $-$ already analyzed in [BR], and called the \it
Maclaurin-Jacobi pattern $-$ \rm which seems to exhibit a second order phase transition (in the sense of
Landau), according to Landau's Thermodynamical Theory.\\\\ In the articles [CMR], [Ho], [In], [Ks], [Mi2],
[Mi3], [MM], [Si], and in the
books [Ly], [SH], we find other formal confirmations of the theoretical result of [BR].\\
Nevertheless, for this pattern seems does not subsist these two criteria (see § 5).\section{The Maclaurin-Jacobi
pattern}The Maclaurin-Jacobi pattern ([Le1]) is a simplified mathematical model of the Self-Gravitating Systems
Theory ([BF], [Ch], [CMR], [FP], [Jar], [Ko], [Lam], [Le1], [Le2], [Lyt], [Pa], [St1], [Ta1], [Ta2]), and of the
Physics of Compact Objects ([KW], [Pa], [ST]).\\\\In Astrophysics, there exists the problem of determining the
(closed) stationary equilibrium configurations of a self-gravitating rotating fluid. If we suppose that the
fluid is subject to a rigid rotation around one of its symmetry axis, under the further simplified hypotheses of
homogeneity, incompressibility and non-viscosity of the rotating fluid, we obtain a first linear solution to
this problem: the Maclaurin-Jacobi pattern.\\ Applied to polytropic fluid configurations with index $n\leq
0.808$ ([Va], [St2]), to the theory of elliptic galaxies ([OP], [St2]) and to the Black Holes thermodynamics
([Da1], [Da2], [Wa]), this model provides theoretical predictions in good agreement with the corresponding
experimental data, so that the Maclaurin-Jacobi pattern is a valid model of the Theoretical
Astrophysics\footnote{Furthermore, there exists satisfactory applications of this model to the Theory of \it
Collective Models \rm of atomic nuclei (see [ALLMNRA], [BK], [CPS1], [CPS2], [IA], [RS], [Ro1],
[Ro2]).}.\\\\Following [CMR], let us consider a homogeneous, incompressible and non-viscous fluid in rigid
rotation around one of its symmetry axes, for example the $Ox_3$ axis of a Cartesian coordinates system
($x_1,x_2,x_3$). We choose a normalized orthogonal reference frame $\{O,\hat{i},\hat{j},\hat{k}\}$, whose origin
$O$ is the center of mass of the rotating fluid. We suppose that a reference frame is rotating with the fluid
(respect to which it is at rest\footnote{In this case, such a reference is properly called \it co-rotating rest
reference frame \rm of the given fluid system.}). \\ If $\vec{\omega}$ is the angular velocity of rotation, let
us suppose that $\vec{\omega}=\omega\hat{k}$, in order that the angular momentum is
$\vec{J}=M\big((\vec{\omega}\times\vec{r})\times\vec{r}\big)$, where $\vec{r}$ is the position vector (measured
in the rest reference frame co-rotating with the fluid) of the generic material point of the system, and $M$ is
the total mass of the fluid. $\vec{J}$ and $\vec{\omega}$ are both parallel to the $Ox_3$ axis.\\\\
Historically, C. Maclaurin ([Mc]), C.G.J. Jacobi ([Jc]) and H. Poincaré ([Po]) dealt with the dynamical problem
of determining the equilibrium configurations of the fluid system. The results obtained by
the first two authors, forms the so-called \it ''Maclaurin-Jacobi model'' \rm(or \it M-J pattern\rm).\\\\
For such a fluid system, in the above mentioned rest reference frame co-rotating with the fluid, the (Euler's)
hydrostatic equilibrium equation is$$\nabla(p(\vec{r})-\varphi(\vec{r}))=0,\leqno(1)$$where $p$ is the
hydrostatic pressure and $\varphi$ the potential of the fluid system. In the rest reference frame, let
$S(\vec{r};J^2)=0$ be the surface equation of the unknown equilibrium configuration. If we suppose
$S(\vec{r};J^2)>0$ inside the fluid, then the region $V_S\subset\mathbb{R}^3$ is bounded, connected and take up
by the fluid, and represented by $S(\vec{r};J^2)\geq 0$. This surface is parametrical dependent by $J^2$, where
$J$ is the modulus of the angular momentum. The potential $\varphi$ of this well-defined self-gravitating fluid,
is given by the sum of the gravitational potential $\varphi_g$ and of the centrifugal potential $\varphi_c$, so
that
$$\varphi[S](\vec{r};J^2)=\varphi_g[S](\vec{r})+\varphi_c[S](\vec{r};J^2),\leqno(2)$$where
$$\varphi_g[S](\vec{r})=\frac{1}{4\pi}\int_{S(\vec{s})\geq 0}\frac{d\vec{s}}{|\vec{r}-\vec{s}|}$$is the
gravitational potential, and$$\varphi_c[S](\vec{r};J^2)=\frac{J^2}{2I^2}(x_1^2+x_2^2),\ \ \
I[S]=\frac{15}{8\pi}\int_{S(\vec{r})\geq 0}(x_1^2+x_2^2)d\vec{r}$$are, respectively, the centrifugal potential
and the inertial moment of the fluid system, computed respect to the axis $Ox_3$, with $\vec{r}=(x_1,x_2,x_3)$.
The gravitational potential satisfies the Poisson's equation$$\Delta\varphi_g[S](\vec{r})=-\rho,\ \
\rho(\vec{r})=1\ \ \mbox{if}\ \ S(\vec{r})\geq 0,\ \ \rho(\vec{r})=0\ \ \mbox{if}\ \ S(\vec{r})<0$$where $\rho$
is the (constant) density of the fluid. Moreover, the associated boundary conditions are the
following$$p(\vec{r}){\big |}_{S(\vec{r})=0}=0,\ \ \ \lim_{r\rightarrow\infty}\varphi_g(\vec{r})=0,\ \ \
\varphi[S](\vec{r};J^2){\big |}_{S(\vec{r})=0}=\mbox{const.},\ \ \ r=|\vec{r}|.$$Through the bifurcation method
(see [CMR], [Pe1], [Pe2], [Dy]), and neglecting, for simplicity, the various stability questions (see [CMR],
[Lyt]), a first (linear) approximated solution of the non-linear differential system
$$\left\{\begin{array}{l}\nabla(p-\varphi)=0\\ \\
\Delta\varphi_g=-\rho\\ \\p(\vec{r}){\big |}_{S(\vec{r})=0}=\lim_{r\rightarrow\infty}\varphi_g[S](\vec{r})=0\\
\\\varphi[S](\vec{r};J^2){\big |}_{S(\vec{r})=0}=\mbox{const.}\ ,\end{array}\right.\leqno(3)$$gives
the so-called \it Maclaurin-Jacobi pattern, \rm that consists of the first two sequences of (infinite and
stationary\footnote{But no static.}) solutions $S(\vec{r};J^2)$ of (3), for suitable values of the
parameter\footnote{Respects to suitable natural units, relative to the physical and geometrical parameters of an
ellipsoid, the parameter $J^2$ takes well-defined values (see [CMR]).} $J^2$. Precisely, we have a first
sequence of (infinite and stationary\footnote{But no static.}) solutions, called \it Maclaurin's sequence \rm
$${\cal S}_{Macl}=\{S(\vec{r};J^2); \ 0<J^2<0,384436\},$$and a second sequence of (infinite and stationary\footnote
{But no static.}) solutions, called \it Jacobi's sequence \rm $${\cal S}_{Jac}=\{S(\vec{r};J^2);\ 0,384436\leq
J^2<0,632243\}.$$The first sequence of solutions is geometrically formed by two-axial ellipsoids (or spheroids)
whose symmetry axis of rotation is $Ox_3$, whereas the second sequence of solutions is geometrically formed by
three-axes ellipsoids. The figures of the first sequence are called \it Maclaurin's ellipsoids \rm(or \it
spheroids\rm), \rm whereas the figures of the second sequence are called \it Jacobi's ellipsoids. \rm When
$\omega\rightarrow 0$, by a corollary of an important theorem of L. Lichtenstein ([Le1], [Li]), the only
stationary and static equilibrium configurations, compatible with (3), are that spherics (corresponding to
$J^2=\omega=0$), so that the relative symmetry group is $O(3)$. Little by little that $\omega$ increases (and
hence $J^2$ increases as well), the fluid mass takes, at first, a two-axial ellipsoidal shape, with symmetry
axis the rotation axis and symmetry group\footnote{As regard the various notions and group notations, let us
follows [CMR], [LL3], Chap. XII., and [Ga], [MZ].} $D_{\infty h}$, whereas, as soon as $J^2$ has values not
lower than 0,384436, the fluid mass takes the shape of a three-axes ellipsoid, with relative symmetry group
$D_{2h}$.\\\\ Increasing $J^2$ beyond the value 0,632243, we obtain another sequence of (infinite and
stationary\footnote{But no static.}) solutions (of the system (3)), represented by new geometrical figures,
called \it pearlike configurations of Poincaré, \rm whose symmetry group is $C_{2v}$. If the latter sequence,
called \it Poincaré's sequence, \rm is
$${\cal S}_{Poin}=\{S(\vec{r};J^2);\ 0,632243\leq J^2<1,346350\},$$then $\{{\cal S}_{Macl},{\cal S}_{Jac},{\cal
S}_{Poin}\}$ will be called \it Maclaurin-Jacobi-Poincaré pattern.\\\\ \rm The interest of this paper is
restricted at discussing the Maclaurin-Jacobi pattern, that includes the first two sequences $\{{\cal
S}_{Macl},{\cal S}_{Jac}\}$ and their temporal evolution ${\cal S}_{Macl}\rightarrow{\cal S}_{Jac}$.
\section{The second order phase transitions}Let us briefly recall
the main definitions of the Landau's phase transition theories.\\\\
$\bullet$ \bf Landau's Thermodynamical Theory. \rm A first semi-empirical theory of phase transitions, was given
by P. Ehrenfest (1933) on the basis of the previous thermodynamical results achieved by J.W. Gibbs and F. Van
der Waals. It is based on the discontinuity (of various orders) of the free energy\footnote{Among the
thermodynamical variables $(P,T,V)$, we chooses $(P,T)$ as independent.} $F(P,T)=U(P,T)-T\cdot S(P,T)$ (with $U$
internal energy, $S$ entropy, $P$ pressure), so that a phase transition is of order $n\geq 1$ if there exists,
at least, a partial derivative of $F$, of order $n$, that exhibits a discontinuity, while all previous partial
derivative of order $m<n$, are continuous. However, this classification was incomplete and not physically consistent.\\\\
In 1937, L.D. Landau ([La]), starting from the results of the predecessors (above mentioned), gave a first
version of a continuum phase transition theory, called \it Landau's thermodynamic theory, \rm (see [LL5], Chap.
XIV, \S 143). This theory, before all, distinguishes between phase transitions 'with order parameter' and phase
transitions 'without order parameter'. The transitions of the first type are those involving phases having the
same symmetry group\footnote{For instance, the two phases of a liquid-gas transition, has $O(3)$ as symmetry
group.} or different symmetry groups not connected by any relation of the type group-subgroup. The transitions
of the second type, instead, regards phases whose symmetry groups are different but related by some relation of
the type group-subgroup. These are characterized (differently from those of the first type) by a symmetry change
of the order parameter, that must be invariant respect to each symmetry groups of the two phases.\\
Besides the symmetry change, the thermodynamical functions, characterizing the physical state of the system
subject to such phase transition, may not assume any values because it must change according to the above
mentioned symmetry change, and in dependence of such order parameter.\\\\ The latter critical analysis have been
systematically developed by Landau in a second time after his first Thermodynamical Theory, and represents the
so-called \it Landau's Phenomenological Theory. \rm Is is stronger (as physical theory) than his initial
Thermodynamical Theory.\\The (possible) counterexample described in this paper, confirms this
non-equivalence.\\\\In the framework of the Thermodynamical Theory, for phase transitions with order parameter
Landau first assumes that the free energy must be function of the order parameter, as an independent and
extensive variable $\eta$, typical of the thermodynamical system under examination\footnote{Examples of order
parameter are the spontaneous polarization in a ferroelectric transition, or the magnetic susceptibility in a
ferromagnetic transition.}. In addition, he assumes (\it Landau's hypothesis \rm) that $F(P,T,\eta)$ must be
analytic\footnote{This is the weak-point of the Landau's theory, since $(P_c,T_c)$ is a simple singularity
where, formally, there is no functional analyticity.} in a neighborhood of the critical point $(P_c,T_c)$ (where
such transition occur), that must be $\eta=0$ in the initial phase ($T<T_c$) and that must be $\eta\neq 0$ in
the final phase ($T>T_c$). Hence, under such conditions, in a neighborhood of the critical point $(P_c,T_c)$, we
must have
$$F(P,T,\eta)=F_0+\alpha(P,T)\eta+A(P,T)\eta^2+B(P,T)\eta^3+C(P,T)\eta^4+....\leqno(4)$$where $F_0=F(P,T,0)$
and $\alpha, A,B,C,...$ are independent of $\eta$. Landau considers only a fourth order expansion of the type
(4) (called \it Landau's polynomial \rm), and assumes\footnote{This symmetry is motivated by theoretical physics
reasons (since we must have the same solution considering both $\eta$ and $-\eta$), and by stability reasons.}
the symmetry $\eta\rightarrow -\eta$, so that, in the expansion (4), the odd degree terms are zero. Moreover,
since we consider equilibrium phase transitions, thermodynamically $F$ should be a minimum in $(P_c,T_c)$ for
$\eta=0$, so that the expansion (4) reduces to the following Landau's polynomial (of fourth degree)
$$F(P,T,\eta)=F_0+A(P,T)\eta^2+C(P,T)\eta^4.\leqno(5)$$In particular, by the minimum necessary conditions for $F$ in
$(P_c,T_c,0)$, we must have$$(\partial F/\partial\eta)_{(P_c,T_c,0)}=0,\ \
(\partial^2F/\partial\eta^2)_{(P_c,T_c,0)}>0,$$that is (via (4))$$A(P,T)=a(P)(T-T_c),\ \
\eta=\eta_0(T-T_c)^{1/2},\ \ C(P_c,T_c)>0,\leqno(6)$$with $\eta_0$ constant.
Landau puts (6) as necessary and sufficient conditions for classify, as continuum, a phase transition.\\ \\
$\bullet$ \bf Landau's Phenomenological Theory \rm (see [LL5], Chap. XIV, \S 145). In Landau's Thermodynamical
Theory (as well as in the previous theories), many important theoretical questions about continuum phase
transitions with order parameter remain unsolved; in this theory subsisted only some vague and unfounded
assumptions. The main question concerns the determination of the functional dependence laws of the values of the
thermodynamical functions on the typical symmetry change of a continuum phase transition with order parameter.
For this purpose, if $G_0\rightsquigarrow G_1$ denotes such a generic phase transition, from the initial phase
with symmetry group $G_0$ to the final phase with symmetry group $G_1$, Landau put $\eta=0$ (and
$G_0$-invariance) in the initial phase, and $\eta\neq 0$ (and $G_1$-invariance) in the final phase. Moreover, he
assumed that the symmetry $G_1$ of the final phase (of the transition), must be the isotropy group of the order
parameter $\eta\neq 0$, respect to a well-defined irreducible representation\footnote{From now on, for
simplicity sake's, if not otherwise specified, any irreducible representation ${\cal D}_{G_0}$ of $G_0$, will be
denoted only by $\cal D$, without any other indices. Furthermore, we suppose that it denotes always an
irreducible representation of $G_0$.} ${\cal D}_{G_0}$ of $G_0$, satisfying further conditions, respectively
called \it Landau's condition \rm and \it Lif\v{s}its's condition \rm (see \S 4).\\ Hence, as regard what has
been said (and that forms the essence of the so-called \it Landau's Phenomenological Theory\rm), it is correct
to denote this transition in the compact notation$^{17}$ $G_0\stackrel{{\cal
D}}{\rightsquigarrow}G_1$.\section{The ''chain subduction criterion'' and the\\''maximality criterion''}Let
$G_0\stackrel{{\cal D}}{\rightsquigarrow}G_1$ be a generic continuum phase transition from an initial phase with
symmetry group $G_0$ (where $\eta=0$ in the pre-existed phase) to the final phase with symmetry group $G_1$
(where $\eta\neq 0$ in the resulting phase), associated to the irreducible representation ${\cal D}$ of $G_0$,
and having order parameter $\eta\in\mathbb{R}^+\cup\{0\}$. In this case, we say (with Landau) that $G_1$ is a
\it permitted \rm subgroup\footnote{This is coherent with the Landau's definition of continuum phase transition
with order parameter, and in agreement to the fact that $G_0$ and $G_1$ must be in
some relation of the type group-subgroup. See, also, the condition 1. that follows.} respect to $\cal D$.\\\\
Subsequently (see [CLP]), according to Landau and other Authors, if this phase transition is of second order,
then the following necessary or sufficient conditions must hold\footnote{The original sources are: for Landau's
condition, see [La], for Lif\v{s}its's condition, see [Lif], for Birman's condition, see [Bi1], and, for
Ascher's condition, see [As1].}:\begin{enumerate}\item $G_1$ is a subgroup of $G_0$,\item \it (Landau's
condition) \rm $([{\cal D}]^3|I(G_0))=0$, that is the symmetric cube of $\cal D$ does not contain the identity
representation $I$ of $G_0$,\item \it(Lif\v{s}its's condition) \rm $(\{{\cal D}\}^2|V(G_0))=0$, that is the
anti-symmetric square of $\cal D$ does not contain any arbitrary vectorial representation $V$ of $G_0$,\item
\it(Birman's subduction criterion) \rm $\cal D$ subduces the identity representation of $G_1$,\item \it
(Ascher's maximality criterion) \rm $G_1$ is a maximal subgroup of $G_0$.\end{enumerate} Successively, F.E.
Goldrich and J.L. Birman ([GB]) introduced, for finite (or countable) groups, a necessary stronger condition
than condition 4. of above ([Bi1]). The integer number\footnote{With $\chi_{{\cal D}}(g)$, we will denotes the
character of the element $g\in G_1$, computed respect to the irreducible representation $\cal D$ of $G_0$,
whereas $|G_1|$ denotes the order of $G_1$.\\ For no finite groups, if we denote the usual Haar's measure on
$G_1$ with $\mu$, then we have $\displaystyle i_{{\cal D}}(G_1)=\int_{G_1}\chi_{{\cal D}}(g)d\mu(g)$, whereas,
for countable groups, by means of the so-called \it Born-Von Kàrmàn cyclicity \rm(\it boundary\rm) \it
conditions, \rm a particular expression of the type (7) is possible (for further details, see [Ja3]).}$$i_{{\cal
D}}(G_1)=\frac{1}{|G_1|}\sum_{g\in G_1}\chi_{{\cal D}}(g),\leqno(7)$$that gives the number of times that ${\cal
D}|_{G_1}$ ($=$restriction of $\cal D$ to $G_1$) contains the identity representation of $G_1$, is called \it
subduction index \rm(or \it frequency\rm) of \ $\cal D$, respect to $G_1$. \rm Therefore, the \it
(Birman-Goldrich) chain subduction criterion \rm ([GB]) states that, if $\cal D$ is an irreducible
representation of $G_0$, $G_1$ and $G_1'$ are subgroups of $G_0$ such that $G_1'$ is subgroup of $G_1$ (that is
$G_1'\subseteq G_1\subseteq G_0$) and $i_{{\cal D}}(G_1)=i_{{\cal D}}(G_1')=1$, then $G_1'$ is not a permitted
subgroup (in accordance with Landau) respect\footnote{In the context of Landau's theory, this means that $G_1'$
cannot be, respect to the irreducible representation $\cal D$, the symmetry group of the final phase of a
continuum phase transition of the type $G_0\stackrel{\cal D}{\rightsquigarrow}G_1'$.} to $\cal D$.\\ In the
consequence of the criticisms moved in [LPC], M.V. Jari\'{c} ([Ja2]) extended this chain subduction criterion to
the more general context including the case where $i_{{\cal D}}(G_1)=i_{{\cal D}}(G_1')$ is any non-negative
integer, reaching to the following, more general \it (Birman-Goldrich-Jari\'{c}) chain subduction criterion\rm.
If $\cal D$ is an irreducible representation of $G_0$, and $G_1'$, $G_1$ are subgroups of $G_0$ such that
$G_1'\subset G_1\subseteq G_0$, with $G_1'$ subgroup of $G_1$ and $i_{{\cal D}}(G_1')=i_{{\cal
D}}(G_1)\in\mathbb{N}\cup\{0\}$, then $G_1'$ is not a (Landau's) permitted subgroup respect to $\cal D$.\\
In the following section, these results will be applied to the Maclaurin-Jacobi pattern.\section{Applications to
Maclaurin-Jacobi pattern} Recalling that G. Bertin and L.A. Radicati (see [BR]) proved that the symmetry change
$D_{\infty h}\rightarrow D_{2h}$, associated to the dynamical evolution\footnote{See the end of § 2.} ${\cal
S}_{Macl}\rightarrow{\cal S}_{Jac}$ of the Maclaurin-Jacobi pattern, may be formally interpreted as a second
order phase transition in the Landau's Thermodynamical Theory.\\ Therefore, this transition should be, also, the
same according to the Landau's Phenomenological Theory (if we assume valid the equivalence between these
theories), so that it may be classified as a transition of the type $D_{\infty h}\stackrel{\cal
D}{\rightsquigarrow}D_{2h}$ respect to some irreducible representation $\cal D$ of $D_{\infty h}$, with
$D_{\infty h}=G_0, D_{2h}=G_1$ and $D_{2h}$ permitted subgroup (of $D_{\infty h}$) respect to $\cal D$.\\
Nevertheless, it is possible to prove that such transition do not respect the chain subduction criterion of
Birman-Goldrich-Jari\'{c}, for any given irreducible representation $\cal D$ of $D_{\infty h}$. In fact, if we
consider the groups $G_1=D_{4h}$, $G_1=D_{6h}$, with $G_1'=D_{2h}$ proper subgroup of both, we have
$D_{2h}\subset D_{4h}\subset D_{\infty h}=G_0$, $D_{2h}\subset D_{6h}\subset D_{\infty h}$, and since we'll see
that $i_{\cal D}(D_{2h})=i_{\cal D}(D_{4h})$, $i_{\cal D}(D_{2h})=i_{\cal D}(D_{6h})$ for any given irreducible
representation $\cal D$ of $D_{\infty h}$, by the above mentioned Birman-Goldrich-Jari\'{c} criterion, it
follows that\footnote{This criterion is applicable to these group chains because $D_{\infty h}$ is countable and
$D_{4h},D_{6h},C_{2v}$ are finite.} $D_{2h}$ is not a (Landau's) permitted subgroup, that is $D_{\infty
h}\stackrel{\cal D}{\rightsquigarrow} D_{2h}$ is not a continuum phase transition (in the sense of Landau's
Phenomenological Theory), for any given irreducible representation $\cal D$.\\\\In the Maclaurin-Jacobi pattern,
we assume the $Oz$ axis as a rotation axis $h$ of the first order. Moreover, let us denote with
$\theta(t)=\omega t$ the angle of the rotation $C_{\theta}$, with $E$ the identity, with $I$ the inversion and
with $C_{2\phi}$ the rotations of $\phi/2$ radians around to the rotation axis of the second order, placed in
the $Oxy$ plane. \\Hence, with these notations and hypotheses, the table of characters of $D_{\infty h}$ is the
follows (see [Co], vol. I; [ITO], App. A,B; [Ja5]):
$$\begin{tabular}{|c|c|c|c|c|c|c|c|}
\hline irr. repr.& $E$ & $C_{\theta},C_{-\theta}$ & $C_{2\phi}$ & $I$ & $IC_{\theta},IC_{-\theta}$ & $IC_{2\phi}$\\
\hline \bfseries $A_{1g}$ & 1 & 1 & 1 & 1 & 1 & 1 \\ \hline \bfseries $A_{1u}$ &1&1&1&-1&-1&-1\\
\hline \bfseries $A_{2g}$ & 1 & 1 & -1 & 1& 1&-1\\ \hline\bfseries $A_{2u}$ &1&1&-1&-1&-1&1\\ \hline\bfseries
$E_{ng}$ &2&2\ $\cos(n\theta$)&0&2&2\ $\cos(n\theta$)&0\\ \hline\bfseries $E_{nu}$ &2&2\
$\cos(n\theta)$&0&-2&-2\ $\cos(n\theta)$&0\\ \hline\bfseries $E_{(n+1/2)g}$
&2&2$\cos(n+1/2)\theta$&0&2&2$\cos(n+1/2)\theta$&0\\ \hline\bfseries
$E_{(n+1/2)u}$&2&2$\cos(n+1/2)\theta$&0&-2&-2$\cos(n+1/2)\theta$&0\\ \hline\end{tabular}$$with $A_{ig},A_{iu}\ \
i=1,2$, real one-dimensional irreducible representations and
$E_{ng},E_{nu},E_{(n+\frac{1}{2})g},E_{(n+\frac{1}{2})u}\ \ n\in\mathbb{N}$, real two-dimensional irreducible
representations. Therefore, considering the group operations $D_{2h},D_{4h},D_{6h}$ (see [Co], vol. I), by (7)
it is immediate to verify that
$$i_{A_{1u}}(D_{6h})=i_{A_{2g}}(D_{6h})=i_{A_{2u}}(D_{6h})=i_{E_{nu}}(D_{6h})=i_{E_{(n+\frac{1}{2})u}}(D_{6h})=0,\ \
\forall n\in\mathbb{N};$$
$$i_{A_{1u}}(D_{4h})=i_{A_{2g}}(D_{4h})=i_{A_{2u}}(D_{4h})=i_{E_{nu}}(D_{4h})=i_{E_{(n+\frac{1}{2})u}}(D_{4h})=0,\
\ \forall n\in\mathbb{N};$$
$$i_{A_{1u}}(D_{2h})=i_{A_{2g}}(D_{2h})=i_{A_{2u}}(D_{2h})=i_{E_{nu}}(D_{2h})=i_{E_{(n+\frac{1}{2})u}}(D_{2h})=0,\ \
\forall n\in\mathbb{N};$$
$$1\leq i_{E_{ng}}(D_{6h})=\frac{1}{6}\big(1+5\cos(n\theta)\big)\leq i_{E_{ng}}(D_{2h})=\frac{1}{2}
\big(1+\cos(n\theta)\big),\ \ \forall n\in\mathbb{N};$$
$$1\leq i_{E_{(n+1/2)g}}(D_{6h})=\frac{1}{6}\big(1+5\cos(n+1/2)\theta\big)\leq$$ $$\leq
i_{E_{(n+1/2)g}}(D_{2h})=\frac{1}{2}\big(1+\cos(n+1/2)\theta\big), \ \ \forall n\in\mathbb{N};$$
$$1\leq i_{E_{ng}}(D_{4h})=\frac{1}{4}\big(1+3\cos(n\theta)\big)\leq i_{E_{ng}}(D_{2h})=\frac{1}{2}
\big(1+\cos(n\theta)\big),\ \ \forall n\in\mathbb{N};$$
$$1\leq i_{E_{(n+1/2)g}}(D_{4h})=\frac{1}{4}(1+3\cos(n+1/2)\theta)\leq$$ $$\leq
i_{E_{(n+1/2)g}}(D_{2h})=\frac{1}{2}(1+\cos(n+{1}/{2})\theta),\ \ \forall n\in\mathbb{N},$$ so that
$$\left.\begin{array}{rl}\big(i_{E_{ng}}(D_{lh})<i_{E_{ng}}(D_{2h})\big)\\ \\
\big(i_{E_{(n+1/2)g}}(D_{lh})<i_{E_{(n+1/2)g}}(D_{2h})\big)\end{array}\right\}
\Leftrightarrow\big(0<\theta<\frac{2k\pi}{n}\big)\ \ \forall n\in\mathbb{N},\ k\in\mathbb{Z}$$ with $l=4,6$,
and, fixing arbitrarily $k$, we have
$$\big(\lim_{n\rightarrow\infty}\frac{2k\pi}{n}=0\big)\Rightarrow\big(\theta(t)=\omega t=0\big),$$hence $\omega=0$
for $t>0$, which is impossible because the case $\omega=0$ corresponds to a static self-gravitating fluid mass
of spherical symmetry $O(3)$ (see §2, and [Le1], [Li]). Therefore
$$i_{E_{ng}}(D_{lh})=i_{E_{ng}}(D_{2h})\geq 1,\ \ i_{E_{(n+1/2)g}}(D_{lh})=i_{E_{(n+1/2)g}}(D_{2h})\geq 1,\ \
\forall n\in\mathbb{N}$$with $l=4,6$. In conclusion, $i_{\cal D}(D_{2h})=i_{\cal D}(D_{lh})\ \ l=4,6$, for any
given irreducible representation $\cal D$ of $D_{\infty h}$, and hence, by the chain subduction criterion,
$D_{\infty h}\rightarrow D_{2h}$ is not a continuum phase transition, in contrast with the conclusion outlined
in [BR]. It follows a restriction to the validity of the chain subduction criterion.\\\\
Finally, the continuum phase transition $D_{\infty h}\rightarrow D_{2h}$ invalidates, also, the Ascher's
maximality criterion because $D_{2h}$ is not a maximal subgroup of $D_{\infty h}$ (hence, also of $O(3)$) since
$D_{2h}\subset D_{lh}\subset D_{\infty h}(\subset O(3))\ \ l=4,6$, whereas, in conformity with such criterium,
$D_{2h}$ should be a maximal subgroup of $D_{\infty h}$.
\section{Prospects and possible applications}The previous critical analysis, leads to a
deeper critical study of the formal setting of the general (macroscopic) phase transition theory, eventually
with the support of the (microscopic) statistical theory of phase transitions, also in connection with the
theory of the first order phase transitions.\\ Indeed, further motivations for the present analysis come from
possible applications of the Maclaurin-Jacobi pattern to particular phase transitions involved in some
astrophysical models of stars\footnote{Besides the possible applications that these arguments could have in the
theory of fragmentation of atomic nuclei, and related phase transitions (see [GMS], [Mig], [RW]).}: for example,
phase transitions may be invoked in the explanation of some aspects of the complex physical phenomenology of the
solid external crust of a radio-pulsar (or of a rotating neutron star, in general), as, for instance, the
(possible) formation of a superficial thin-layer of gases if we have a first order phase transition instead of a
second order one.\\On the other hand, a similar question is already connected with the problem of the supernova
explosions (see [Gr], vol. II, Chap.XVIII, §9, sect. a)), where (following W.A. Fowler and F. Hoyle) a
particular phase transition - neutrons producing - occurs, involving a photodisintegration of $^{56}Fe$-atoms,
forming a (solid) crystalline lattice imbedded in a degenerate (and strongly anisotropic - if there is in action
an intense magnetic field\footnote{The magnetic field configurations (see [Kc], [OPF]) of a star, are in
relation with the occurrence of a phase transitions of the first order or of the second order, with different
physical implications in both cases.}) Coulomb electron gas (see [Gr], [KW], [Pa], [ST], [Ta2]).\\Again, a
similar question enters into the problem of the determination of the shapes left by a nova remnants from a
rotating white dwarf parent (see [FH]).\section{References}{[ALLMNRA]} G. Andersson, S.E. Larsson, G. Leader, P.
M\"{o}ller, S.G. Nilsson, I. Ragnarsson, S, \AA berg, \it Nuclear Shell Structure at very high Angular Momentum,
\rm Nuclear Physics, A 268 (1976) 205-256.\\{[As1]} E. Ascher, Physics Letters, 20 (1966) 352-354.\\{[As2]} E.
Ascher, \it Permutation Representations, Epikernels and Phase Transitions, \rm Journal of Physics C: Solid State
Physics, 10 (1977) 1365-1377.\\{[AK]} E. Ascher, J. Kobayashi, \it Symmetry and Phase Transitions: the Inverse
Landau Problem, \rm Journal of Physics C: Solid State Physics, 10 (1977) 1349-1363.\\{[BK]} R. Beringer, W.J.
Knox, \it Liquid-Drop Nuclear Model with High Angular Momentum, \rm Physical Review, n. 4, 121 (1961)
1195-1200.\\ {[BR]} G. Bertin, L.A. Radicati, \it The Bifurcation from the Maclaurin to the Jacobi Sequences as
a Second-Order Phase Transition, \rm The Astrophysical Journal, 206 (1976) 815-821.\\{[BF]} B. Bertotti, P.
Farinella, \it Physics of the Solar System, \rm Kluwer Academic Publishers, Dordrecht, 2003.\\ {[Bi1]} J.L.
Birman, \it Simplified Theory of Symmetry Change in Second-Order Phase Transitions: application to $V_3Si$, \rm
Physical Review Letters, 17 (1966) 1216-1219.\\{[Bi2]} J.L. Birman, \it Group Theory of the Landau-Thermodynamic
Theory of Continuous Phase Transitions in Crystals, \rm in: \it Group Theoretical Methods in Physics, \rm
Lecture Notes in Physics, vol. 79, ed. by P. Kramer, A. Rieckers, Springer-Verlag, Berlin, 1978.\\{[Bo]} N.
Boccara, \it Symétries Brisées, \rm Hermann, Paris, 1976.\\{[Ch]} S. Chandrasekhar, \it Ellipsoidal Figures of
Equilibrium, \rm Yale University Press, New Haven and London, 1969.\\{[CPS1]} S. Cohen, F. Plasil, J.W.
Swiatecki, \it Equilibrium shapes of a rotating charged drop and consequences for heavy ion induced nuclear
reactions, \rm in: \it Proceedings of the third Conference on Reactions between Complex Nuclei, \rm A. Ghiorso,
R.M. Diamond and H.E. Conzett Eds., University of California Press, 1963.\\{[CPS2]} S. Cohen, F. Plasil, W.J.
Swiatecki, \it Equilibrium Configurations of Rotating Charged or Gravitating Liquid Masses with Surface Tension,
\rm Annals of Physics, 82 (1974) 557-596.\\{[CMR]} D.H. Constantinescu, L. Michel, L.A. Radicati, \it
Spontaneous Symmetry Breaking and Bifurcations from the Maclaurin and Jacobi Ellipsoids, \rm Le Journal de
Physique, 40 (1979) 147-159.\\{[Co]} J.F. Cornwell, \it Group Theory in Physics, \rm 3 voll., Academic Press,
London, 1984.\\{[CLP]} A.P. Cracknell, J. Lorenc, J.A. Przystawa, \it Landau's Theory of Second-Order Phase
Transitions and its Application to Ferromagnetism, \rm Journal of Physics C: Solid State Physics, 9 (1976)
1731-1758.\\{[Da1]} P.C.W. Davies, \it The Thermodynamic Theory of Black Holes, \rm Proceedings of the Royal
Society of London, Series A, 353 (1977) 499-521.\\{[Da2]} P.C.W. Davies, \it Thermodynamics of Black Holes, \rm
Reports on Progress in Physics, 41 (1978) 1313-1355.\\{[Dy]} F. Dyson, \it Dynamics of a Spinning Gas Cloud, \rm
Journal of Mathematics and Mechanics, 18 (1968) 91-101.\\{[FJ]} R.L. Fiedler, T.W. Jones, \it White Dwarf
rotation and the shapes of nova remnants, \rm The Astrophysical Journal, 239 (1980) 253-256.\\{[FH]} W.A.
Fowler, F. Hoyle, \it Neutrino Processes and Pair Formation in Massive Stars and Supernovae, \rm The
Astrophysical Journal, 9 (1964) 201-319.\\{[FP]} A.M. Friedman, V.L. Polyachenko, \it Physics of Gravitating
Systems, \rm voll. I,II, Springer-Verlag, New York, 1984.\\{[GL]} V. Gabis, M. Lagache (Eds.), \it Les
Transformations de Phases dans les Solides Minéraux, \rm 2 Voll., Société Fran\c{c}aise de Minéralogie et de
Cristallographie, Paris, 1981.\\{[Ga]} G. Gaeta, \it Bifurcation and Symmetry Breaking, \rm Physics Reports, 189
(1990) 1-87.\\ {[GB]} F.E. Goldrich, J.L. Birman, \it Theory of Symmetry Change in Second-Order Phase
Transitions in Perovskite Structure, \rm The Physical Review, 167 (1968) 528-532.\\{[Gr]} L. Gratton, \it
Introduzione all'Astrofisica (Stelle e Galassie), \rm 2 voll., Zani-\\chelli Editore, Bologna, 1977.\\{[GRS]}
D.H.E. Gross, M.E. Madjet, O. Scapiro, \it Fragmentation phase transition in atomic cluster I, II, III, IV, \rm
Zeitschrift f\"{u}r Physik, D 39 (1997) 75-83, 309-316, D 41 (1997) 219-227, B 104 (1997) 541-551.\\{[Ho]} M.
Hosoya, \it Improvement of the Landau Theory of Phase Transitions, \rm Journal of the Physical Society of Japan,
42 (1977) 399-407.\\{[IA]} F. Iachello, A. Arima, \it The Interacting Boson Model, \rm Cambridge University
Press, Cambridge, 1987.\\ {[In]} V.L. Indenbom, \it Phase Transitions without Change in the Number of Atoms in
the Unit Cell of the Crystal, \rm Soviet Physics-Crystallography, 5 (1960) 106-115.\\{[ITO]} T. Inui, Y. Tanabe,
Y. Onodera, \it Group Theory and Its Application in Physics, \rm Springer-Verlag, Berlin 1990.\\{[Jc]} C.G.J.
Jacobi, \it \"{U}ber die Figur des Gleichgewichts, \rm Poggendorf Annalen der Physik und Chemie, 3 (1834)
229-238.\\{[Jar]} W.S. Jardetzky, \it Theories of Figures of Celestial Bodies, \rm Interscience Publishers Inc.,
New York, 1958.\\{[Ja1]} M.V. Jari\'{c}, \it Group Theory and Phase Transitions, \rm Physica, 114A (1982)
550-556.\\{[Ja2]} M.V. Jari\'{c}, \it Spontaneous Symmetry Breaking and the Chain Criterion, \rm The Physical
Review B, 23 (1981) 3460-3463.\\{[Ja3]} M.V. Jari\'{c}, \it Structural Phase Transitions in Crystals:
Broken-Symmetry (Isotropy) Groups, \rm Journal of Mathematical Physics, 24 (1983) 2865-2882.\\{[Ja4]} M.V.
Jari\'{c}, \it Landau Theory, Symmetry Breaking and the Chain Criterion, \rm in: \it Group Theoretical Methods
in Physics, \rm Lecture Notes in Physics, vol. 135, ed. by K.B. Wolf, Springer-Verlag, Berlin, 1980.\\{[Ja5]}
M.V. Jari\'{c}, \it How to Calculate Isotropy Subgroups of a Crystallographic Space Group, \rm in: \it Group
Theoretical Methods in Physics, \rm ed. by W.W. Zachary, World Scientific Publishing Company, Singapore,
1986.\\{[Ja6]} M.V. Jari\'{c}, \it Counterexamples to the Maximality Conjecture of Landau-Higgs Models, \rm in:
\it Group Theoretical Methods in Physics, \rm Lecture Notes in Physics, vol. 201, ed. by G. Denardo,
Springer-Verlag, Berlin, 1984.\\{[Ja7]} M.V. Jari\'{c}, \it Landau Theory, Symmetry Breaking and the Chain
Criterion, \rm in: \it Group Theoretical Methods in Physics, \rm Lecture Notes in Physics, vol. 135, ed. by K.B.
Wolf, Springer-Verlag, Berlin, 1980.\\{[JB]} M.V. Jari\'{c}, J.L. Birman, \it Group Theory of Phase Transitions
in A-15 $O^3_h$-Pm3n Structure, \rm The Physical Review B, 16 (1977) 2564-2568.\\{[KW]} S.W. Koch, \it Dynamics
of First-Order Phase Transition in Equilibrium and Nonequilibrium Systems, \rm Lecture Notes in Physics n. 207,
Springer-Verlag, Berlin, 1984.\\{[Ko]} Z. Kopal, \it Figures of Equilibrium of Celestial Bodies, \rm The
University of Wisconsin Press, Madison, 1960.\\{[Ks]} V.A. Koptsik, \it Polymorphic Phase Transitions and
Symmetry, \rm Soviet Physics-Crystallography, 5 (1961) 889-898.\\{[Lam]} H. Lamb, \it Hydrodynamics, \rm
(1$^{th}$ edition 1932) 6$^{th}$ edition, Cambridge University Press, Cambridge, 1957-1974.\\{[La]} L.D. Landau,
\it Zur Theorie der Phasenumwandlungen I, II, \rm Zh. \'{E}ksp. Teor. Fiz., 7 (1937) 19, 627; French trans.,
Phys. Z. Sowjetunion, 11 (1937) 26-47, 545-555, and reprinted in: \it Collected Papers of L.D. Landau, \rm ed.
by D. Ter Haar, Pergamon Press, London, 1962.\\{[LL3]} L.D. Landau, E.M. Lif\v{s}its, \it Fisica Teorica 3:
Meccanica Quantistica (Teoria Non Relativistica), \rm Editori Riuniti-Edizioni Mir, Roma, 1982.\\{[LL5]} L.D.
Landau, E.M. Lif\v{s}its, \it Fisica Teorica 5: Fisica Statistica, \rm Editori Riuniti-Edizioni Mir, Roma,
1978.\\{[Le1]} N.R. Lebovitz, \it Rotating Fluid Masses, \rm Annual Review of Astronomy and Astrophysics, 5
(1967) 465-480.\\{[Le2]} N.R. Lebovitz, \it Bifurcation and Stability Problems in Astrophysics, \rm in: \it
Applications of Bifurcation Theory, \rm ed. by P.H. Rabinowitz, Academic Press, New York, 1977.\\{[Li]} L.
Lichtenstein, \it Gleichgewichtsfiguren Rotierender Fl\"{u}ssigkeiten, \rm Verlag Von Julius Springer, Berlin,
1933.\\ {[Lif]} E.M. Lif\v{s}its, Zh. \'{E}ksp. Teor. Fiz., 11 (1941) 255, Soviet Journal of Physics, 6 (1942)
61-74, 252-263.\\{[LPC]} J. Lorenc, J. Przystawa, A.P. Cracknell, \it A Comment on the Chain Subduction
Criterion, \rm Journal of Physics C: Solid State Physics, 13 (1980) 1955-1961.\\{[Lyt]} R.A. Lyttleton, \it The
Stability of Rotating Fluid Masses, \rm Cambridge University Press, Cambridge, 1953.\\{[Ly]} G.Ya. Lyubarskij,
\it The Application of Group Theory in Physics, \rm Pergamon Press, London, 1960.\\{[Mc]} C. Maclaurin, \it
Treatise of Fluxions, \rm Edinburgh, 1742.\\{[Mi1]} L. Michel, \it Symmetry Defects and Broken Symmetry.
Configurations Hidden Symmetry, \rm Reviews of Modern Physics, 52 (1980) 617-651.\\{[Mi2]} L. Michel, \it Les
Brisures Spontanées de Symétrie en Physique, \rm Le Journal de Physique, Colloque C7, Supplément au n. 11, Tome
36 (1975) page C7-41.\\{[Mi3]} L. Michel, \it Minima of Higgs-Landau Polynomials, \rm CERN Report n. TH-2716
(1979), reprinted in: \it Regards sur la Physique Contemporaine, \rm ed. by H. Bacry, Editions CNRS, Paris,
1979.\\{[MM]} L. Michel, J. Mozrymas, \it Application of Morse Theory to the Symmetry Breaking in the Landau
Theory of Second Order Phase Transition, \rm in: \it Group Theoretical Methods in Physics, \rm Lecture Notes in
Physics, vol. 79, ed. by P. Kramer, A. Rieckers, Springer-Verlag, Berlin, 1983.\\{[MZ]} L. Michel, B.I.
Zhilinskiì, \it Symmetry, Invariants, Topology I: Basic Tools, \rm Physics Reports, 341 (2001) 11-84.\\{[Mig]}
A.B. Migdal, \it Phase transition in nuclear matter and non-pair nuclear forces, \rm Soviet Physics JETP, n. 6,
36 (1973) 1052-1055.\\{[OP]} J.P.O. Ostriker, P.J.E. Peebles, \it A numerical study of the stability of flattned
galaxies: or, can cold galaxies survive?, \rm The Astrophysical Journal, 186 (1973) 467-480.\\{[OPF]} F.J.
Owens, C.P. Poole, H.A. Farach (Eds.), \it Magnetic Resonance of Phase Transitions, \rm Academic Press, New
York, 1979.\\ {[Pa]} T. Padmanabhan, \it Theoretical Astrophysics, \rm 3 voll., Cambridge University Press,
Cambridge, 2000.\\{[Pe1]} J.M. Perdang, \it Irregular Stellar Variability, \rm in: \it Chaos in Astrophysics,
\rm NATO-ASI Series C, vol. 161, ed. by J.R. Buchler, J.M. Perdang and E.A. Spiegel, D. Reidel Publishing
Company, Dordrecht, 1985.\\{[Pe2]} J.M. Perdang, \it On Some Group-Theoretical Aspects of the Study of
Non-Radial Oscillations, \rm Astrophysical Space Sciences, 18 (1968) 355-371.\\{[PP]} G.A. Pik-Pichak, \it
Equilibrium shapes and Fission of a rotating nucleus, \rm Soviet Physics JETP, 16 (5) (1963) 1201-1206.\\{[Po]}
H. Poincaré, \it Sur l'équilibre d'une masse fluide animée d'un mouvement de rotation, \rm  Acta Math., 7 (1885)
259-380.\\{[RW]} J. Richert, P. Wagner, \it Microscopic model approaches to fragmentation of nuclei and phase
transitions in nuclear matter, \rm Physics Reports, 350 (2001) 1-92.\\{[RS]} P. Ring, P. Schuck, \it The nuclear
many-body problem, \rm Springer-Verlag, New York, 1980.\\{[Ro1]} C.E. Rosenkilde, \it Stability of Axisymmetric
Figures of Equilibrium of a Rotating Charged Liquid Drop, \rm Journal of Mathematical Physics, n. 1, 8 (1967)
98-118.\\{[Ro2]} C.E. Rosenkilde, \it Surface-Energy Tensors, \rm Journal of Mathematical Physics, n. 1, 8
(1967) 84-97.\\{[ST]} S.L. Shapiro, S.A. Teukolsky, \it Black Holes, White Dwarfs and Neutron Stars (The Physics
of Compact Objects), \rm John Wiley and Sons, New York, 1983.\\{[Si]} Yu.I. Sirotin, \it Possible Changes in the
Point-Group Magnetic Symmetry in Second-Order Ferromagnetic Transitions, \rm Soviet Physics-Crystallography, 8
(1963) 195-196.\\{[St1]} M. Stiavelli, \it Modelli Dinamici di Galassie Ellittiche, \rm Tesi di Laurea, Facoltà
di Scienze MM.FF.NN. dell'Università degli Studi di Pisa, Corso di Laurea in Fisica, Anno Accademico
1982-1983.\\{[St2]} M. Stiavelli, \it Symmetry and Symmetry Breaking in Astrophysics, \rm in: \it Symmetry in
Nature, a volume in honour of Luigi A. Radicati di Br\'{o}zolo, \rm 2 voll., Pubblications of the Scuola Normale
Superiore, Pisa, 1988.\\{[SH]} H.T. Stokes, D.M. Hatch, \it Isotropy Subgroups of the 230 Crystallogra-\\phic
Space Groups, \rm World Scientific Publishing Company, Singapore, 1988.\\ {[SA]} M. Sutton, R.L. Armstrong, \it
Symmetry Restrictions on Phase Transitions Imposed by Group-Subgroup Structure, \rm The Physical Review B, 25
(1982) 1813-1821.\\{[Ta1]} J.L. Tassoul, \it Theory of Rotating Stars, \rm Princeton University Press,
Princeton, New Jersey, 1978.\\{[Ta2]} J.L. Tassoul, \it Stellar Rotation, \rm Cambridge University Press,
Cambridge, 2000.\\{[Th]} M. Tinkham, \it Group Theory and Quantum Mechanics, \rm McGraw-Hill Book Company, New
York 1964.\\{[To1]} J.C. Tolédano, P. Tolédano, \it The Landau Theory of Phase Transitions, \rm World Scientific
Publishing Company, Singapore, 1987.\\{[To2]} J.C. Tolédano, P. Tolédano, \it A Counterexample to the <<Maximal
Subgroup Rule>> for Continuous Crystalline Transitions, \rm Le Journal de Physique, 41 (1980) 189-192.\\{[Va]}
P.O. Vandervoort, \it The Equilibrium of a Galactic Bar, \rm The Astrophysical Journal, 240 (1980)
478-487.\\{[Wa]} R.M. Wald, \it Quantum Field Theory in curved space-time and Black Hole Thermodynamics, \rm The
University of Chicago Press Ltd., Chicago and London, 1994.

\end{document}